\documentclass[sigconf]{acmart}

\AtBeginDocument{%
  }

\usepackage{subfigure}
\usepackage{multirow}
\usepackage{tabularx}
\usepackage{makecell}
\pdfoutput=1

\renewcommand\footnotetextcopyrightpermission[1]{}
\setcopyright{none}
\renewcommand\footnotetextcopyrightpermission[1]{}
\setcopyright{none}
\makeatletter
\renewcommand\@formatdoi[1]{\ignorespaces}
\makeatother

\setcopyright{none} 
\makeatletter
\def\@copyrightspace{\relax}
\makeatother
\begin{document}

\title{Advancing Network Security: A Comprehensive Testbed and Dataset for Machine Learning-Based Intrusion Detection}

\author{Talaya Farasat}

\affiliation{%
	\institution{University of Passau}
	\streetaddress{Passau, Germany}
	\city{Passau}
	\country{Germany}}
	
\author{JongWon Kim}

\affiliation{%
	\institution{GIST}
	\streetaddress{Gwangju, South Korea}
	\city{Gwangju}
	\country{South Korea}}

\author{Joachim Posegga}

\affiliation{%
	\institution{University of Passau}
	\streetaddress{Passau, Germany}
	\city{Passau}
	\country{Germany}}

\renewcommand{\shortauthors}{Farasat et al.}

\begin{abstract}
 This paper introduces a Testbed designed for generating network traffic, leveraging the capabilities of containers, Kubernetes, and eBPF/XDP technologies. Our Testbed serves as an advanced platform for producing network traffic for machine learning based network experiments. By utilizing this Testbed, we offer small malicious network traffic dataset publically that satisfy ground truth property completely.
\end{abstract}




\maketitle
\pagestyle{plain}

\section{Introduction}

Existing network intrusion datasets are typically collected in a static manner, making them difficult to expand and unable to guarantee a perfect ground truth. These limitations are fundamentally tied to the design of traditional Testbeds \cite{traffic}. To address these issues, new datasets need to be built from scratch on Testbeds that incorporate the latest revolutionary technologies. By transitioning from virtual machines to containers, we can generate scalable, modular, and heterogeneous network traffic. Each container, being highly specialized for its purpose, enables the creation of datasets with perfect ground truth, a goal that has been unattainable in previous network traffic datasets \cite{traffic}.

Cloud-native computing is an efficient strategy for managing diversified workloads on a shared cluster using containers \cite{jongsu}. Kubernetes [7], an open-source container orchestration engine, automates the management and deployment of containerized applications. Linux eBPF is a light-weight small virtual machine that provides a set of libraries which allows dynamic injection of codes from user space into various kernel events. Meanwhile, XDP provides special hooks for eBPF kernel programs to rapidly pass, drop/filter, and redirect network packets received at networking ports \cite{talia}.

In this paper, we present a Testbed for network traffic generation that leverages Docker containers, Kubernetes, and eBPF/XDP. Utilizing this Testbed, we publicly release malicious network traffic that fully satisfies the ground truth property. Additionally, the Testbed includes GPU-enabled (TESLA T4) ML processing machine for machine learning processing, allowing researchers to conduct machine learning-based experiments.

\section{Testbed Architecture}
Aligned with future Internet Testbeds like the Global Environment for Networking Innovation (GENI), the Gwangju Institute of Science and Technology (GIST) has launched the OF@TEIN Testbed/playground \cite{oftein, aris,aris2, usman, talia, talia2}. The OF@TEIN playground provides a dynamic array of distributed physical and virtual resources for users and developers to learn operational and development issues and conduct various experiments. It is an overlay, multi-site playground spanning heterogeneous underlay networks, connecting around 14 sites across 10 countries. The playground features "SmartX Micro-Boxes," commodity server-based hyper-converged resources (compute/storage/networking) distributed at multi-site edge locations. These Micro-Boxes support experiments around Cloud computing, Software-defined Networking (SDN), and Network Function Virtualization (NFV). Equipped with cloud-native (containerized) functionalities, the Micro-Boxes function as Kubernetes-orchestrated workers with SDN-coordinated connectivity to other Micro-Boxes. The infrastructure is managed by the "SmartX Playground Tower," which automates the building, operation, and utilization of the playground. The Playground Tower oversees the multi-site playground operation, leveraging several entities, including a Visibility Center, to ensure efficient management.

To set up our Testbed, we request two switches and seven physical machines from the OF@TEIN playground administration. The entire setup is located at GIST and interconnected through the OF@TEIN playground network, as illustrated in Figure 1. The hardware specifications for the machines are detailed in Table 1. We utilize Docker for container virtualization and Kubernetes (with the Weave Net plugin) for container orchestration. Additionally, we employ eBPF/XDP for network traffic capturing. Our testbed configuration includes:
one Master Node and four Worker Nodes dedicated to network threat simulations, one storage machine designated for dataset storage, and one ML processing machine for running machine learning algorithms. This setup allows researcher to efficiently simulate and analyze network threats, store relevant data, and process machine learning tasks, thereby providing a comprehensive environment for research and development activities.

\subsection {\textbf{Malicious Traffic Generation}}
We generate malicious traffic by simulating various network threats in our Testbed. This traffic is collected using eBPF/XDP from the virtual container interfaces (veth-pairs) within our Testbed. All network traffic is then transferred to a storage machine using the Python Paramiko library. Subsequently, we can move the necessary data to the ML-processing machine to conduct machine learning-based experiments.

\begin{figure}
	\centering
	\includegraphics[width=80mm, height=6.0cm]{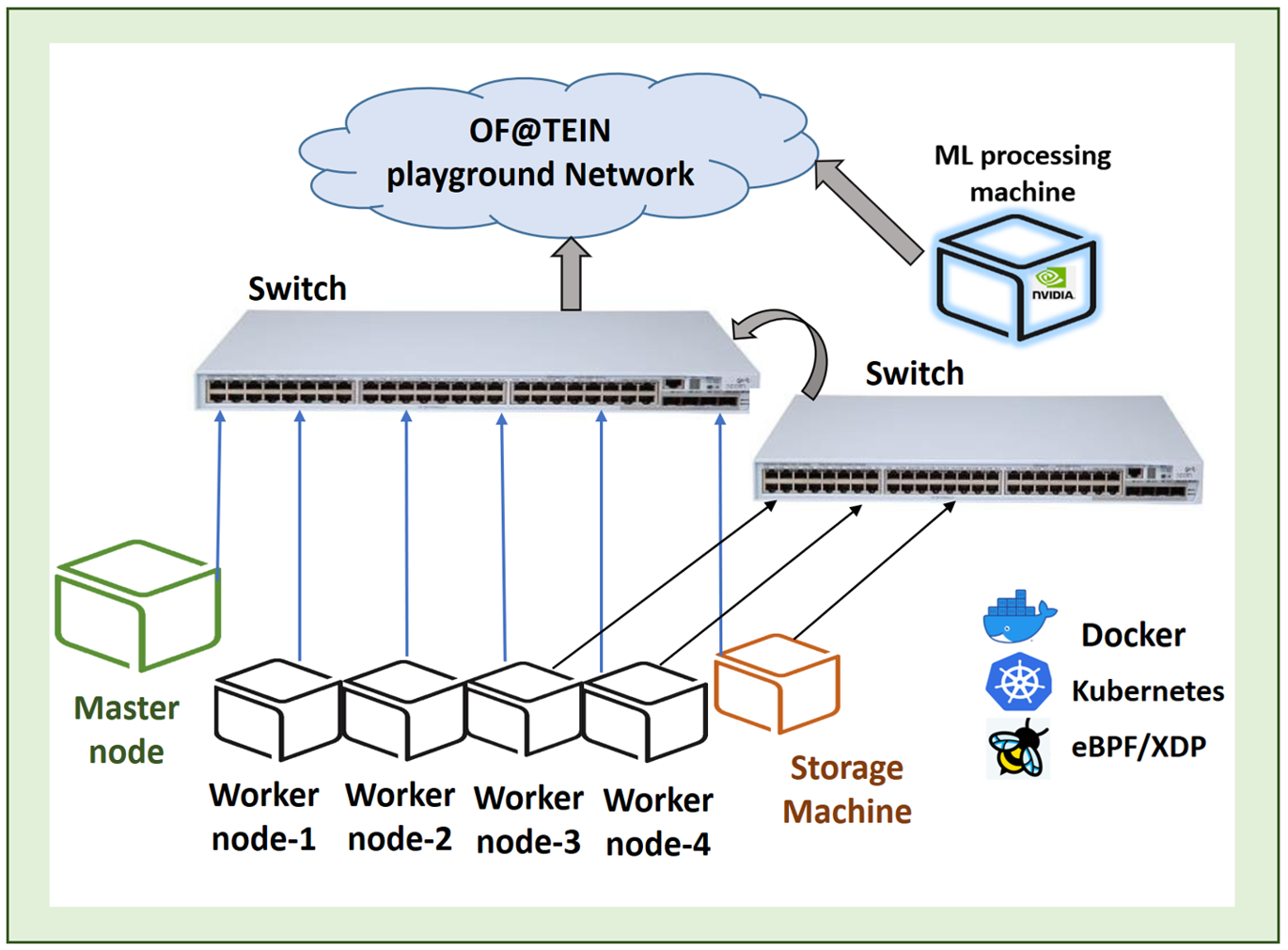}
	\caption{Testbed setup for malicious traffic generation (All Resources obtained from OF@TEIN playground)}
\end{figure}

\begin{table}	
		\renewcommand{\arraystretch}{0.7}
		\begin{tabular}{|p{1.7cm}|c|c|c|}
			\hline
			{\textbf{Hardware}} & {\textbf{Specifications}}\\
			
			\hline
			Switches & \makecell{Netgear M4300-52G-PoE+} \\ \hline
			Master and Worker Nodes& \makecell{Intel® Xeon® CPU D-1518 \\ Model: SuperServer E300-8D}\\ \hline
			Storage Machine& \makecell{Intel®Xeon CPU E5-2690 V2@3.00GHz, \\ Model: Server, Memory DDR3 12x8GB, HDD 5.5TB}  \\ \hline
			ML Processing Machine & \makecell{Intel® Xeon® CPU D-2183IT, \\ Model: SuperServer E403-9D, \\ configured NVIDIA Tesla T4} \\ \hline		
			
		\end{tabular}
		\caption{Hardware Specifications of Testbed for malicious traffic generation}
		
	\end{table}

\begin{table} 
	\renewcommand{\arraystretch}{1}

	\begin{tabular}{|p{1.4cm}|c|c|c|c|c|}
		\hline
		\cline{1-3}
		{\textbf{Attacks}} & {\textbf{Tools}} & {\textbf{Attack Types}}\\
		\cline{1-3}
		 
		\hline{}
		DoS& hping3 & TCP SYN flood \\ \cline{2-3}
		& HULK  & DoS Get  \\ \cline{1-3}
		DDoS& hping3 & \makecell{TCP SYN flood, \\ ICMP flood,\\  TCP sequence prediction} \\ \cline{2-3}
		& Slowhttptest & \makecell{Slowloris, \\ Slow-body,\\ Slow-read, \\ Slow-range} \\ \cline{1-3}
		Brute Force& Hydra  & \makecell{Credentials \\ hijacking (username, \\ password)} \\ \cline{1-3}
		
		Heartbleed& Metasploit Framework & Bug in OpenSSL  \\ \cline{1-3}

	\end{tabular}
	\caption{Types of Attacks that are performed on our Testbed}
\end{table}

\subsubsection{\textbf{Attack Scenerios:}}
To create malicious network traffic, we conduct various types of attacks on our Testbed. Figures 2, 3, and 4 illustrate the attack scenarios, while Table 2 lists the attack types implemented. Each attack is executed individually, with a duration of approximately one minute, within February 2022. We use the Kubernetes node selector property to assign specific Pods to designated nodes.

\begin{figure*}[tp]
	\centering
	\includegraphics[width=130mm, height=9cm]{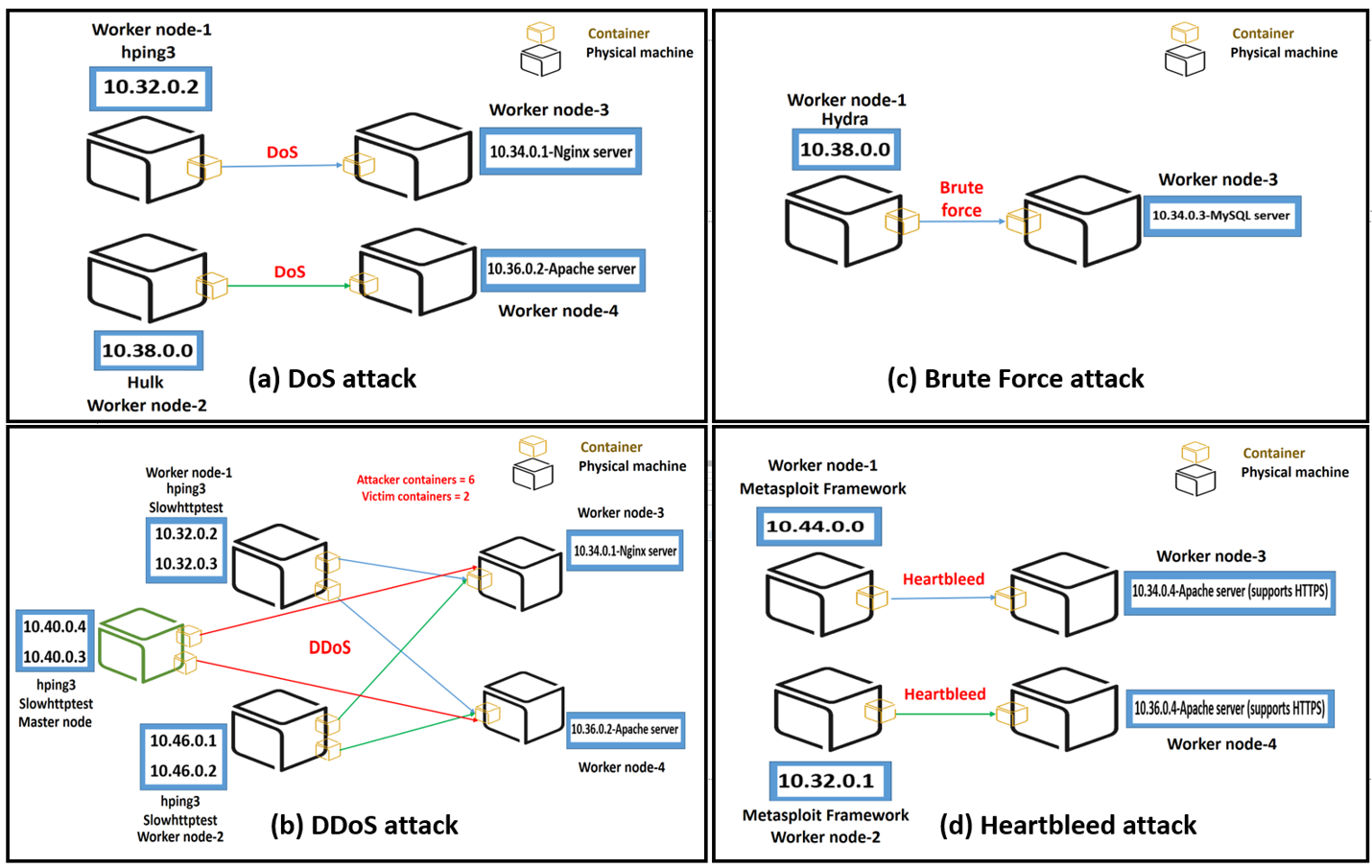}
	\caption{Attack Scenerios}
\end{figure*}

\begin{figure}[tp]
	\centering
	\includegraphics[width=90mm, height=5.2cm]{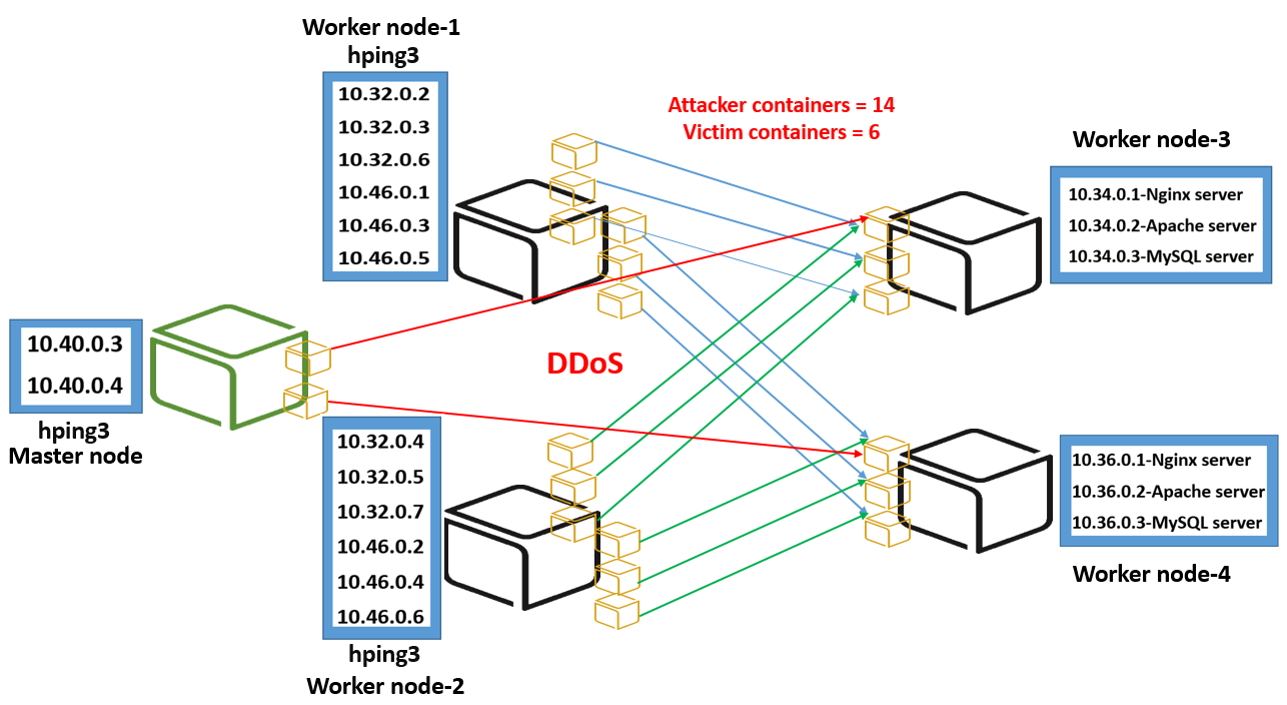}
	\caption{Attack Scenerios (Large DDoS attack (14 attackers and 6 victims))}
\end{figure}

\begin{figure}[tp]
	\centering
	\includegraphics[width=90mm, height=5.2cm]{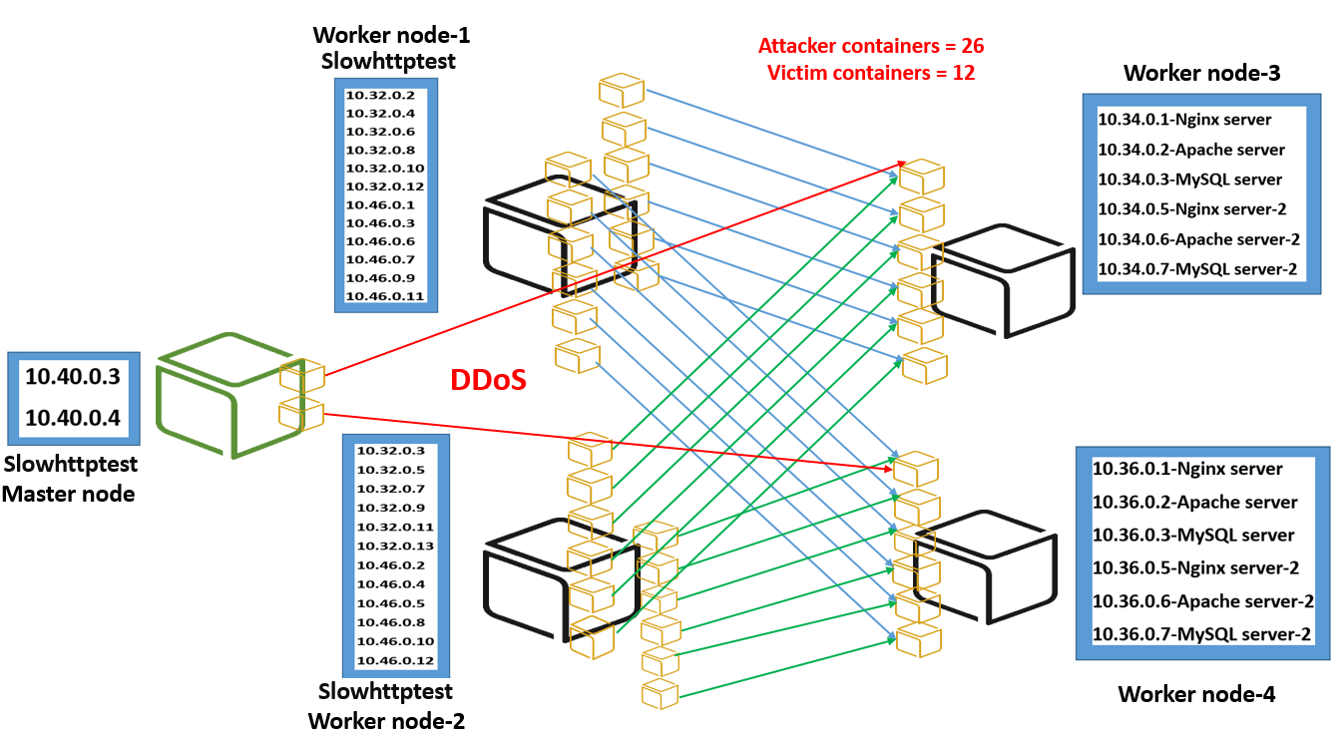}
	\caption{Attack Scenerios (Large DDoS attack (26 attackers and 12 victims))}
\end{figure}

\begin{figure*}[tp]
	\centering
	\includegraphics[width=110mm, height=4.0cm]{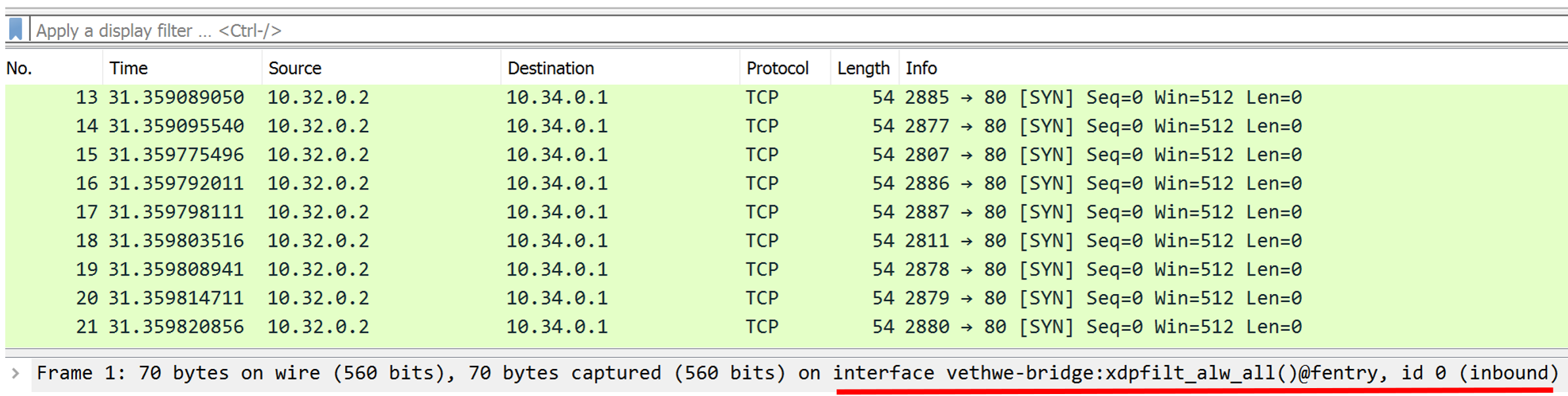}
	\caption{DDoS .pcap file from interface vethwe-bridge}
\end{figure*}
\begin{figure*}[tp]
	\centering
	\includegraphics[width=110mm, height=5.0cm]{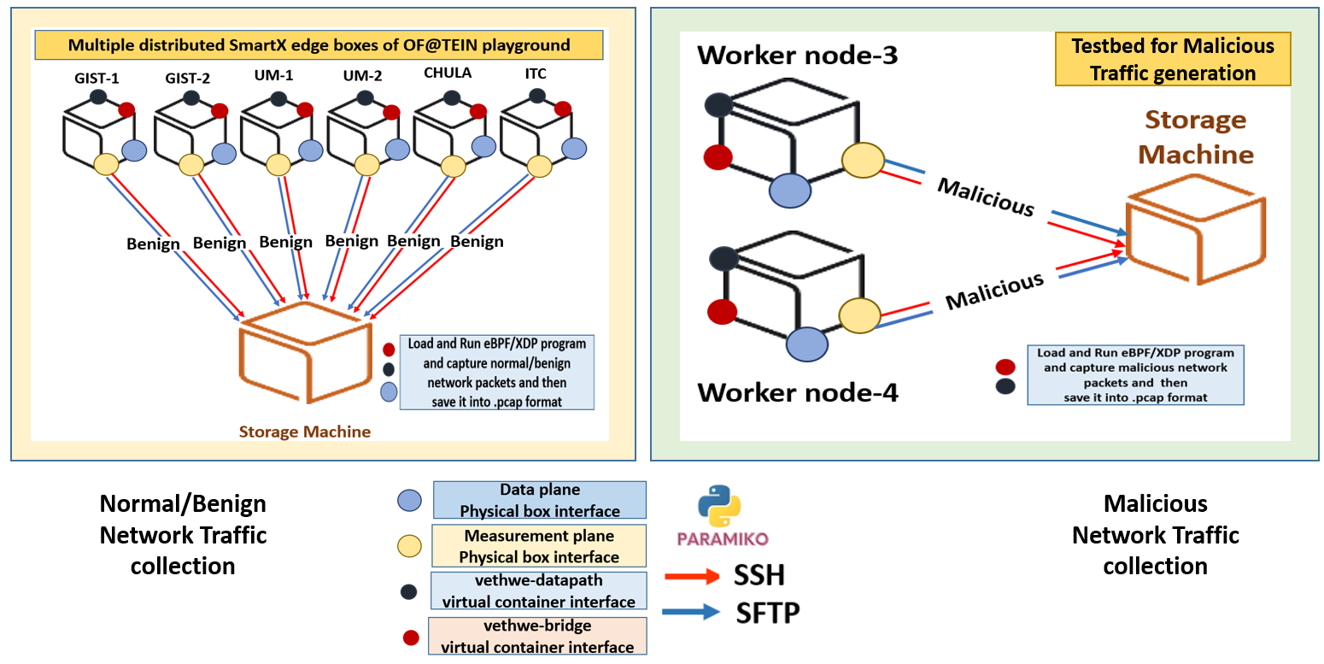}
	\caption{Collection of Dataset}
\end{figure*}

\textbf{DoS Attacks:} We perform two DoS attacks with the docker images of hping3 (TCP SYN flood attack) \cite{hping} and HULK (GET attack) \cite{hulk}. We use Nginx server \cite{nginx} and Apache server docker images \cite{apache} as victims. In our DoS attack setup, we designate Worker Node 1 and Worker Node 2 as attackers, and Worker Node 3 and Worker Node 4 as victims. Specifically, the hping3 Pod is deployed on Worker Node 1, and the Nginx server Pod is deployed on Worker Node 3. The HULK Pod is deployed on Worker Node 2, and the Apache server Pod is deployed on Worker Node 4.

In the TCP SYN flood scenario, the container running hping3 on Worker Node 1 attacks the container running the Nginx server on Worker Node 3. Similarly, in the DoS GET attack scenario, the container running HULK on Worker Node 2 attacks the container running the Apache server on Worker Node 4, as illustrated in Figure 4(a).

\textbf{DDoS Attacks:} We perform seven types of DDoS attacks with the docker images of hping3 (TCP SYN flood attack, ICMP flood attack, and TCP sequence prediction attack) and Slowhttptest (Slowloris, Slow-read, Slow-body, and Slow-range) \cite{slow}. We use Nginx server and Apache server docker images as a victims. n DDoS attacks, we utilize a Master Node, Worker Node 1, and Worker Node 2 as attackers, while Worker Node 3 and Worker Node 4 serve as victims. We specify the Pods of hping3 and Slowhttptest on Master Node, Worker Node 1, and Worker Node 2. Pods regarding victims such as Nginx Server and Apache server are deployed on Worker Node 3 and Worker Node 4. 

In the case of the Slowloris attack, containers (Slowhttptest) deployed on the Master node, Worker Node 1, and Worker Node 2 simultaneously attack the containers (Nginx server and Apache server) deployed on Worker Node 3 and Worker Node 4, as illustrated in Figure 2(b). Specifically, each attacker node hosts two containers: one targets the Nginx server container on Worker Node 3, and the other targets the Apache server container on Worker Node 4. We perform all DDoS attacks in this manner except for Slow-range attack, for which, we only use Apache Server containers as a victim on both Worker Node 3 (with IP Address: 10.34.0.2) and Worker Node 4.

\textbf{Brute Force Attack:} We perform Brute Force attack with the docker image of Hydra \cite{hydra}. We use MySQL server \cite{mysql} docker image as victim which requires credentials to use the database. In Brute Force attacks, we utilize Worker Node 1 as attacker and Worker Node 3 as victim. We specify the Pod of Hydra on Worker Node 1 and Pod of MySQL server on Worker Node 3. In this case, container (Hydra) deployed on Worker Node 1 targets on the container (MySQL server) deployed on Worker Node 3, see Figure 2(c). 

\textbf{Heartbleed Attack:} We perform a Heartbleed attack with the Metasploit framework \cite{meta}. We use a Heartbleed (docker image) \cite{heart}, in which Apache server (supports HTTPS traffic) is used as a victim. In Heartbleed attacks, we use Worker Node 1 and Worker Node 2 as attackers and Worker Node 3 and Worker Node 4 as victims. In the first case, Metasploit framework configured on Worker Node 1 targets the container (Apache server) deployed on Worker Node 3. In the second case, Metasploit framework configured on Worker Node 2 targets the container (Apache server) deployed on Worker Node 4, see Figure 2(d).

\textbf{Large DDoS Attacks:} Because this Testbed is based on containerization, it is highly scalable. This allows for the execution of attacks with a large number of attacker and victim containers, enabling extensive experiments. We perform two scenarios of large DDoS attacks as illustrated in Figure 3 and 4. In these attacks, we use Master Node, Worker Node 1, and Worker Node 2 as attackers and Worker Node 3 and Worker Node 4 as victims. 

In first case, we perform DDoS attack with the docker image of hping3 (TCP SYN flood attack). We specify the Pods of hping3 on Master Node, Worker Node 1, and Worker Node 2. Pods regarding victims such as Nginx Server, Apache server, and MySQL server are deployed on Worker Node 3 and Worker Node 4. We deploy 14 attacker containers (hping3) and six victim containers (Nginx Server, Apache server, and MySQL server). So, basically, 3 attacker containers (hping3) from Worker Node 1 attack on 3 victim containers (Nginx Server, Apache server, and MySQL server) on Worker Node 3, and 3 attacker containers (hping3) from Worker Node 1 attack on 3 victim containers (Nginx Server, Apache server, and MYSQL server) on Worker Node 4. The same is the case with Worker Node 2. 3 attacker containers (hping3) from Worker Node 2 attack on 3 victim containers (Nginx Server, Apache server, and MySQL server) on Worker Node 3 and 3 attacker containers (hping3) from Worker Node 2 attack on 3 victim containers (Nginx Server, Apache server, and MYSQL server) on Worker Node 4. From Master Node one attacker container targets one victim container (Nginx Server) on Worker Node 3 and one attacker container targets one victim container (Nginx Server) on Worker Node 4, see Figure 3.

In the second case, we perform a DDoS attack with the docker image of Slowhttptest (Slowloris attack). We deploy 26 attacker containers (Slowhttptest) and 12 victim containers (Nginx Server, Apache server, and MySQL server). We specify the Pods of Slowhttptest on Master Node, Worker Node 1, and Worker Node 2. Pods regarding victims such as Nginx Server, Apache server, and MySQL server are deployed on Worker Node 3 and Worker Node 4. So, basically, 6 attacker containers (Slowhttptest) from Worker Node 1 attack on 6 victim containers (Nginx Server, Apache server, and MySQL server, Nginx Server-2, Apache server-2, and MySQL server-2) on Worker Node 3, and 6 attacker containers (Slowhttptest) from Worker Node 1 attack on 6 victim containers (Nginx Server, Apache server, MySQL server, Nginx Server-2, Apache server-2, and MySQL server-2) on Worker Node 4. The same is the case with Worker Node 2. 6 attacker containers (Slowhttptest) from Worker Node 2 attack on 6 victim containers (Nginx Server, Apache server, MySQL server, Nginx Server-2, Apache server-2, and MySQL server-2) on Worker Node 3, and 6 attacker containers (Slowhttptest) from Worker Node 2 attack on 6 victim containers (Nginx Server, Apache server, MySQL server, Nginx Server-2, Apache server-2, and MySQL server-2) on Worker Node 4. From Master Node one attacker container targets one victim container (Nginx Server) on Worker Node 3 and one attacker container targets one victim container (Nginx Server) on Worker Node 4, see Figure 4.

\section{Dataset Collection}
\textbf{Collection of Malicious Network Data:}
To collect the malicious network traffic dataset, we configure and load an eBPF/XDP program (using native mode) on the interfaces that handle container traffic, specifically the veth-pairs of Worker Node 3 and Worker Node 4. The eBPF/XDP program is loaded on two key interfaces: vethwe-datapath and vethwe-bridge, to capture both incoming and outgoing malicious network traffic. We highlight interface information in the Figure 7 .pcap file. To facilitate analysis, we capture incoming and outgoing network traffic (in .pcap format)  during attacks in separate .pcap files, ensuring clear distinction and organization of the dataset.

We transfer the malicious traffic files from Worker Node 3 and Worker Node 4 to our storage machine using the Python Paramiko library, as illustrated in Figure 6. To prevent congestion and ensure efficient data transfer, we utilize a dedicated data plane. This data plane is configured with a separate VLAN and IP subnet on a NETGEAR switch, isolating the transfer process from the main network traffic.

\textbf{Collection of Normal Network Data:}
Although normal traffic generation is not the primary focus of this paper, we have collected some normal traffic data from the OF@TEIN playground. This normal network data collection is intended to facilitate research experiments and provide a small comprehensive dataset for research analysis.

We gather normal network traffic from various distributed SmartX edge Micro-boxes within the OF@TEIN playground, including GIST-1 (South Korea), GIST-2 (South Korea), UM-1 (Malaysia), UM-2 (Malaysia), CHULA (Thailand), and ITC (Cambodia), as depicted in Figure 6. Our collection targets both virtual container interfaces (veth pairs) and physical interfaces across these distributed SmartX edge boxes. To accomplish this, we configure eBPF/XDP modules on each distributed SmartX edge box. Additionally, we develop a Python script to automate the capture and storage of normal/benign network packets in .pcap format. This script triggers the capture process every minute, ensuring continuous collection of network packets. Utilizing the Python Paramiko library, we seamlessly transfer these .pcap files to a designated storage machine via the data plane. This automated process guarantees uninterrupted collection and storage of normal/benign traffic data. Our data collection efforts specifically target the period of February 2022, providing a comprehensive dataset for subsequent research and analysis.

	\section{Conclusion}
	
In this research, we present a specialized Testbed designed for the generation of network intrusion detection datasets, leveraging containers, Kubernetes, and eBPF/XDP technologies. While in this work our primary focus is on making malicious network traffic, our versatile Testbed is also capable of generating normal network traffic. Furthermore, we extend our dataset collection by gathering a small quantity of normal network traffic from the OF@TEIN playground. We believe that this Testbed and dataset serves as a valuable resource for conducting network intrusion detection experiments, particularly those grounded in machine learning methodologies. We provide our prepared dataset publically and can be accessed here:
https://drive.google.com/drive/folders/11DDvarZg9YOvzPfK2e78CLP7S\_a9bCXm
	\section{Acknowledgement}
This work was partly supported by Institute of Information \& Communications Technology Planning \& Evaluation (IITP) grant funded by the Korea government (MSIT) (No.2019-0-01842, Artificial Intelligence Graduate School Program (GIST)).


\bibliographystyle{ACM-Reference-Format}
\bibliography{samplebase}



\end{document}